\documentstyle[preprint,aps]{revtex}
\input epsf.tex
\def\DESepsf(#1 width #2){\epsfxsize=#2 \epsfbox{#1}}
\begin{document}

\preprint{\vbox{\hbox{OITS-602}\hbox{UCRHEP-T165}\hbox{}}}
\draft
\title {Electric dipole moments and b-$\tau$ unification in the presence of an
intermediate scale in SUSY grand unification}
\author{\bf N. G. Deshpande $^{\dagger} $, B. Dutta$^{\dagger} $, and E.  Keith
$^{\ast} $ }\address{$^{\dagger} $ Institute of Theoretical Science, University
of Oregon, Eugene, OR 97403\\$^{\ast} $ Department of Physics, University of
California, Riverside, CA 92521}
\date{May, 1996}
\maketitle
\begin{abstract}  We show that an intermediate gauge symmetry breaking scale can
be a significant source of electric dipole moments for the electron and neutron
in supersymmetric grand unified theories. New  phases, similar to that of the
CKM matrix, appear which  do not arise from the  supersymmetry (SUSY) breaking
operators. To illustrate, we choose some grand unified SUSY models having an
intermediate gauge symmetry breaking scale with some attractive features. We
also show how well the
$b-\tau$ unification hypothesis works in this class of models.
\end{abstract}
\newpage Supersymmetric grand unified theories (GUTs) with intermediate gauge
symmetry breaking scales are attractive because they resolve a few longstanding
problems and possess some desirable phenomenological features. For example, in
models where the intermediate breaking scale
$M_I \sim 10^{10}-10^{12}$ GeV, one can naturally  get a neutrino mass in the
interesting range of
$\sim 3-10$ eV, which could serve as hot dark matter to explain the observed
large scale structure formation of the universe \cite{[SS]}. The window
$\sim 10^{10}-10^{12}$ GeV is also of the right size for a hypothetical
PQ-symmetry to be broken so as to solve the strong CP problem without creating
phenomenological or cosmological problems \cite{[axion]}. Models which allow
even lower intermediate gauge symmetry breaking scale e.g $M_I\sim 1$ TeV are
also interesting since they predict relatively light new gauge fields, as for
example SU(2)$_R$ charged gauge bosons
$W_R$. In all these intermediate scale models,  lepton flavor violation
 is predicted \cite{[ddk]} which may be close to the current experimental
limit, and hence could provide a signal of such models.

In this letter, we point out another special feature of these intermediate scale
models: they can give rise to detectable amounts of electric dipole moments
(EDMs) to the  electron and neutron. This feature does not depend on the nature
of the ultimate unifying group. We will always assume that supersymmetry is
broken via soft breaking terms introduced at a super high scale. We shall
assume that the    soft breaking terms at the high
scale at which they are introduced are flavor blind  and  CP invariant. It
has already  been shown
\cite{[dh],[BHS],[AS]} for  SUSY SO(10) models without
an intermediate scale that  there could be significant amounts of EDM for
the electron and neutron. However for this to arise,  the universal boundary
condition for the soft SUSY breaking terms has to be implemented at a scale
higher than the GUT scale
$M_G$ such as the reduced Planck or string scales. Consequently, the EDMs for
the electron and neutron are not  expected to be produced in such a manner in
SUSY GUTs with the attractive feature of gauge unification taking place at the
string scale.  The models with an intermediate gauge symmetry breaking scale
however give rise to electron and neutron EDMs,  irrespective of whether a
universal soft SUSY breaking  boundary condition appears at the GUT scale or
above it.  With the boundary conditions we have chosen, we accurately
calculate
the intermediate scale effects on the EDMs. We also discuss the
$b-\tau$ unification hypothesis for SUSY GUTs with an  intermediate gauge
symmetry breaking scale. Whenever in SUSY SO(10) grand unification without an
intermediate scale a tau neutrino mass is desired in the interesting eV range,
it is found that
$b-\tau$ unification hypotheses has to be abandoned \cite{[ML10]}. With the
introduction of the intermediate scale, we examine how well that hypothesis
works
for the various models considered here.

We know that the intermediate scale gauge symmetry breaking theories with
SU(2)$_L\times$SU(2)$_R\times$SU(4)$_C$ \cite{[ddku]} or SU(2)$_L\times$SU(2)$_R
\times$U(1)$_{B-L}\times$SU(3)$_C$ \cite{[ddk]} as the intermediate scale
gauge group, give rise to large lepton flavor violation which  could be
detected through processes like
$\mu\rightarrow e\gamma$. The reason is quite simple.  With
 intermediate gauge symmetry SU(2)$_L\times$SU(2)$_R\times$SU(4)$_C$, the quarks
and leptons are unified. Hence, the $\tau$-neutrino Yukawa coupling is the same
as the top Yukawa coupling. Through the renormalization group equations (RGEs),
the effect of the large
$\tau$-neutrino Yukawa coupling is to make the third generation sleptons lighter
than the first two generations, thus mitigating the GIM cancellation in one-loop
leptonic flavor changing processes involving virtual sleptons. Although the
quarks and leptons are not unified beneath the GUT scale when the intermediate
scale gauge group is SU(2)$_L\times$SU(2)$_R\times$U(1)$_{B-L}\times$SU(3)$_c$,
the same effect is produced from the assumption that the top quark
 Yukawa coupling is equal to the $\tau$-neutrino Yukawa coupling at the GUT
scale.

When we calculate the EDM of the electron the  above stated  principle applies,
but we must also consider the phases at the  gaugino-slepton-lepton vertices.
Likewise, to generate the EDM for the neutron one needs the third
generation down
squark to be lighter than those of the other two generations, which occurs
due to
the large top Yukawa coupling, and  new phases at the gaugino-squark-quark
vertices. In fact, whenever there is  an intermediate scale, irrespective of the
intermediate gauge group (e.g. SU(2)$_L\times$SU(2)$_R\times$SU(4)$_C$ or
SU(2)$_L\times$SU(2)$_R
\times$U(1)$_{B-L}\times$SU(3)$_C$)  such phases  are generated. The reason for
this is that right-handed quarks or leptons are unified in a multiplet in a
given
generation. The superpotential for an intermediate gauge symmetry breaking model
can be written (in the case of SU(2)$_L\times$SU(2)$_R\times$SU(4)$_C$ ):
\begin{eqnarray} W_Y&=&{\bf \lambda_{F_u}}{ F}{ \Phi_2}{ {\bar F}} +
 {\bf\lambda_{F_d}}{ F}{ \Phi_1}{{\bar F}} \, ,
\end{eqnarray}  where $F$ and ${\bar F}$ are the superfields containing
the standard model fermion fields and transform as $(2,1,4)$ and $(1,2,{\bar
4})$ respectively and we have suppressed the generation and gauge group
indices. We choose to work in a basis where
${\bf
\lambda_{F_u}}$ is diagonal in which $W_Y$ can be expressed as the following:
\begin{eqnarray} W_Y&=&{ F}{\bf {\bar\lambda_{F_u}}}{ {\bar F}}{
\Phi_2} +
 { F}{\bf {\bf U^*}{\bar\lambda_{F_d}}}{\bf U^{\dag}}{{\bar F}}{ \Phi_1}\, .
\end{eqnarray} The matrix {\bf U} is a general
$3\times3$ unitary matrix with 3 angles and 6 phases. It can be written as
follows:
\begin{eqnarray} {\bf U}={\bf S^{\prime *}} {\bf V} {\bf S}\, ,
\end{eqnarray} where {\bf V} is the CKM matrix and ${\bf S}$ and ${\bf
S^\prime}$ are  diagonal phase matrices. At the scale
$M_I$ the superpotential for the Yukawa coupling can be expressed in the
following manner:
\begin{eqnarray} W_{\rm MSSM}&=&Q{\bf \bar\lambda_u}{\bf U^c}H_2
+Q{\bf V^*}{\bf \bar\lambda_d}{\bf S}^2{\bf V^{\dag}}D^cH_1+ +E^c{\bf
V_I^*}{\bf
\bar\lambda_L}{\bf S}^2{\bf V_I^{\dag}}LH_1\, ,
 \end{eqnarray}
where the ability to reduce the number of phases by redefinition of fields has
been taken advantage of to the fullest extent possible,
 \begin{eqnarray}
{\bf S}^2\equiv \left(\matrix{
               {\rm e}^{i\phi_d}                &0      &0     \cr
               0      &{\rm e}^{i\phi_s}              &0\cr
               0                &0        &0   }\right)
\end{eqnarray}
 is a diagonal phase matrix with two independent phases, and ${\bf V_I}$ is the
CKM matrix at the intermediate gauge symmetry breaking scale. It is not possible
to do a superfield rotation on
$D^c$ or
$L$ to remove the right handed angle since at $M_I$ the third diagonal element
of the scalar mass matrices
${\bf m_D^2}$ and
${\bf m_L^2}$ develop  differently from the other two diagonal
elements due to  large top Yukawa coupling RGE effects.
When the intermediate gauge symmetry group is
SU(2)$_L\times$SU(2)$_R\times$U(1)$_{B-L}\times$SU(3)$_C$,  the
additional CKM-like phases will be generated in exactly the same way  as
described above.

 The expressions from which we calculate EDMs are given below \cite{[dh]}.\\ For
the electron's EDM we have:
\begin{eqnarray}
\left|d_e\right|=e\left|F_2\right| \left|V_{I_{td}}/V_{I_{ts}}\right|
\left| \sin\phi \right|\, ,
\end{eqnarray} where $F_2$ is given by
\begin{equation}\begin{array}{ll} F_2 =
\displaystyle\frac{\alpha}{4\pi\cos^2\theta_{\rm W}}m_\tau V_{\tau\mu}^{\rm e}
V_{\tau e}^{\rm e}(V_{\tau\tau}^{\rm e*})^2
(A_e+\mu\tan\beta)\times\\[3mm]
\phantom{F_2 =}\times [G_2(m_{\tilde{\tau}_L}^2,m_{\tilde{\tau}_R}^2)-
G_2(m_{\tilde{e}_L}^2,m_{\tilde{\tau}_R}^2)-
G_2(m_{\tilde{\tau}_L}^2,m_{\tilde{e}_R}^2)+
G_2(m_{\tilde{e}_L}^2,m_{\tilde{e}_R}^2)]\, ,
\end{array}
\end{equation} and $V^e_{ab}$ are the matrix elements  of the matrix $V_I$ and
the functions
$G_2(a,b)$ are defined in Eqn. (20) in Ref.
\cite{[BHS]}. $\phi$ includes effects of all possible phases. For the neutron:
\begin{eqnarray}
\left|d_n\right|=4/9 e\left|F_2^\prime\right| \left| \sin\phi \right|
\end{eqnarray}  where $F_2^\prime$ is given by .
\begin{equation}\begin{array}{ll} F_2^\prime =
\displaystyle\frac{\alpha_s}{4 \pi} V_{td} V_{td}(V_{tb}^{*})^2
(A_d+\mu\tan\beta)\times\\[3mm]
\phantom{F_2 =}\times [G_2(m_{\tilde{b}_L}^2,m_{\tilde{b}_R}^2)-
G_2(m_{\tilde{d}_L}^2,m_{\tilde{b}_R}^2)-
G_2(m_{\tilde{b}_L}^2,m_{\tilde{d}_R}^2)+
G_2(m_{\tilde{d}_L}^2,m_{\tilde{d}_R}^2)]\, .
\end{array}
\end{equation} To calculate the squark, slepton and gaugino masses at the low
scale we numerically run the RGEs from the GUT scale down to weak scale. We
assume a universal boundary condition at the GUT scale $M_G$ i.e. all gaugino
masses
$M_i(M_G) =m_{1\over 2}$, all tri-linear scalar couplings $A_i(M_G)= A_0$, and
all soft scalar masses
$m^2_i (M_G)=m_0^2$. We use the RGEs for a given intermediate gauge group from
$M_G$ down to $M_I$ as given in Ref.
\cite{[ddk],[ddku]}. From $M_I$ scale down to the weak scale, we use the
MSSM RGEs
(see, for example, Ref. \cite{[SUSYYUK]}).

As examples to illustrate our point, we choose to  use the following  four
 intermediate scale models:  \\Model (1): It is based on the gauge group
SU(2)$_R\times$SU(2)$_L\times$SU(4)$_C$
\cite{[ddku]}with gauge couplings that are found to be   unified at a scale
$M_G$  near the string unification scale.
   The model breaks to the minimal  supersymmetric standard model at a scale
$M_I\sim 10^{12}$ GeV and can have both large and small $\tan\beta$ scenarios.
 For high $\tan\beta$ scenario we use $M_I= 10^{12}$ GeV and
$M_G\approx 10^{18.26}$ GeV leading to $\alpha_s(M_Z)\approx 0.126$ and for low
$\tan\beta$ scenario we use $M_I= 10^{12}$ GeV and
$M_G\approx 10^{17.83}$ GeV leading to $\alpha_s(M_Z)\approx 0.119$.\\Model (2):
It is Case V of Ref.
\cite{[MohaLee]} where SO(10) is broken down to
 SU(2)$_L\times$SU(2)$_R\times$U(1)$_{B-L}\times$SU(3)$_C$ gauge symmetry at
the scale $M_G$. In this scenario, we use
$M_I\approx 10^{12}$ GeV and
$M_G\approx 10^{15.6}$ GeV leading to $\alpha_s(M_Z)\approx 0.129$.\\ Model
(3):This is the model presented in Ref. \cite{[EMa]} with SO(10) breaking  down
to
 SU(2)$_L\times$SU(2)$_R\times$U(1)$_{B-L}\times$SU(3)$_C$. It is the only
example we use for which D-parity is not broken at $M_G$ and hence left-right
parity ($g_L=g_R$) is preserved in $G_I$.  The field content
allows
$M_I\sim 1$ TeV with
$M_G\approx 10^{16}$ GeV. We use MSSM below the scale $M_I$ for convenience
although in the original work \cite{[EMa]} the two Higgs doublet model  has
been used. The value of
$\alpha_s$ with the MSSM below $M_I$ is about  0.129.\\ Model (4):This is the
model discussed in Ref.\cite{[DKR]}. Once again, SO(10) is broken down to
 SU(2)$_L\times$SU(2)$_R\times$U(1)$_{B-L}\times$SU(3)$_C$ at $M_G$. In this
model,
$M_G$ is predicted to be exactly the same as in the conventional SUSY SO(10)
breaking with no intermediate scale and the scale $M_I$ can have any value
between the TeV and the GUT scales.  Since
there is only one Higgs bidoublet, this model prefers large values
of
$\tan{\beta}$ with $\lambda_t =\lambda_b$ at $M_I$. Nevertheless, the
introduction of nonrenormalizable operators can allow for small $\tan\beta$.

In the Figs. 1(a)-1(d), we plot
\begin{eqnarray}d_r\equiv {\rm Log}_{10}{\left( {d_e/{\rm
sin}\phi\over 4.3\cdot 10^{-27}}\right)}\, ,\end{eqnarray}
where $d_e$ is the EDM of electron, as
a function of the scalar mass
$m_0$ for different values of $m_{1/2}$. Since the EDM of neutron is also of the
same order, we do not plot them. Also, experimentally the EDM of electron
is more
constrained. The experimental bounds are given as :
$d_n<0.8\cdot 10^{-25} $ecm \cite{[dn]} and $d_e<4.3\cdot 10^{-27}$
ecm\cite{[de]}. From the graphs it appears that one can use EDM as a signal for
the intermediate scale in a grand unification scenario. The value of
$\left| {\rm  sin}\phi\right|$ could have arbitrary values from 0 to 1, but
there is no reason to expect it to  be suppressed.

Now we discuss  the viability of $b-\tau$ unification hypothesis in
the  models we have discussed.  The value
for
$m_b^{\rm pole}$ from the existing data is
$m_b^{\rm pole}$=4.75$\pm$ .05 calculated in Ref. \cite{[mb]}, and we use
$m_\tau =1.777$ GeV. The predicted $m_b$ mass in these
intermediate scale models mainly depend on 3 factors: the value of
$\lambda_{t_{G}}$,
$\alpha_s (M_Z)$ and the location of the intermediate scale $M_I$. Using larger
values of
$\lambda_{t_{G}}$ of course lowers the $m_b$ mass, while using larger values
of
$\alpha_s$ increases it. For these models at the scale
$M_G$,  we have used the maximum perturbative value for the top Yukawa coupling
which is about  3.54. For  model (1), since leptons and down quarks are
unified in the same multiplet at the intermediate scales we have
$\lambda_b=\lambda_\tau\neq \lambda_t=\lambda_{\nu_{\tau}}$ for the low
$\tan\beta$ scenario. We find $m_b^{\rm pole}=4.78$ GeV. For the large
$\tan\beta$ version of that model, we have
$\lambda_t=\lambda_b=\lambda_\tau=\lambda_{\nu_{\tau}}$ instead, and find
$m_b^{\rm pole}=4.80$ GeV. We find that model (1) is able to provide  very
reasonable b-quark mass predictions since $\alpha_s$ is of  moderate values
and since down to the scale $M_I$ the relation $\lambda_b =\lambda_\tau$
exists intact.  For models (2) and (3) with low values for
$\tan\beta$, we have
$\lambda_t=\lambda_{\nu_{\tau}}$ and
$\lambda_b=\lambda_\tau$ only at the GUT scale and find $m_b$ pole masses of
$5.76$ GeV and $6.20$ GeV, respectively. However if we had used smaller values
of $\alpha_s$ as used in the original references \cite{[MohaLee],[EMa]}for
those models or had we assumed large values for $\tan\beta$, these masses would
be much closer to the desired range. As in Ref. \cite{[BM]}, one could
purposefully construct models with $M_I\sim 10^{12}$ GeV and lower values for
$\alpha_s$ so as to improve the b-quark mass prediction.  For model (4) with
$\lambda_t=\lambda_b=\lambda_\tau=\lambda_{\nu_{\tau}}$ at $M_G$, in Fig. 2(a)
we choose to  plot the
$m_b^{\rm pole}$ mass as a function of
$M_I$ since the intermediate gauge symmetry breaking  scale in that model
can lie anywhere between the weak scale and the GUT scale. Notice that the
b-quark mass at first increases as the intermediate gauge symmetry breaking
scale  moves away from the GUT scale. But, it then reaches a peak value when the
intermediate scale is about
$10^8$ GeV, and then for  $M_I$ less than that scale it decreases. The
reason for this behavior can be found in the RGEs for  $\lambda_b$ and
$\lambda_\tau$. The RGE for $\lambda_b$ feels the influence of the large top
Yukawa coupling while
$\lambda_\tau$ instead feels the influence of the
$\tau$ neutrino coupling. Though the magnitude of the top and the $\tau$
neutrino couplings are same at the GUT scale, the $\tau$-neutrino coupling
decreases faster than the top Yukawa coupling and reaches its fixed point
sooner. If the Intermediate breaking scale is decreased  $\lambda_b (M_I)$
would also decrease, however  $\lambda_\tau (M_I)$ would not decreases as much.
So, effectively the mass of $m_b$ decreases, since
$m_b$ mass depends on the ratio of $\lambda_b$ to $\lambda_\tau$. We further
note that the interesting values for the intermediate gauge symmetry breaking
scale
$M_I\sim 1$ TeV and
$M_I\sim 10^{12}$ GeV can both give good values for the b-quark mass. Effects of this low
intermediate scale could be observed in the future colliders. In Fig.
2(b), we assume the possibility of model (4) allowing a range of values for
$\tan\beta$ in order to plot the $m_b^{\rm pole}$ as a function of
$\tan{\beta}$ for the interesting case of $M_I=10^{12}$ GeV. We see that
larger values of $\tan\beta$ are preferred and give very reasonable values for
the b-mass. In both Figs. 2(a) and 2(b), we show results for two different
values of $\alpha_s$ as explained in the figure caption.

In conclusion, we find that intermediate gauge symmetry breaking can be a
significant source of electric dipole moments for the electron and neutron. We
have  illustrated this effect for four different models. One of which has the
gauge couplings  unified at the string scale and the others at the usual GUT
scale ($\sim 10^{16}$ GeV). In all the models,  the universal SUSY soft breaking
boundary condition is assumed to be introduced at the GUT scale $M_G$.   Of
course for the models where the gauge unification scale is of order 10$^{16}$
GeV, if we had assumed the boundary condition at the reduced Planck scale or
string scale the EDM predictions would have been further enhanced.  We also
examined how well the
$b-\tau $ unification hypothesis works for these models, and find that
sometimes it works very well especially when the intermediate gauge symmetry
unifies quarks and leptons or when $\tan\beta$ is large.

We thank E. Ma for very useful discussion. This work was supported by Department
of Energy grants DE-FG06-854ER 40224 and DE-FG02-94ER 40837.

\newpage

\leftline{{\Large\bf Figure captions}}
\begin{itemize}
\item[Fig. 1(a)~:] {$d_r\equiv {\rm Log}_{10}{\left( {d_e/\left(
4.27\cdot 10^{-27}{\rm sin}\phi\right)}\right)}$ for the low $\tan\beta$ version of model (1) is plotted as a function of of the universal soft SUSY breaking
mass $m_0$.\\ The solid lines correspond to
$\mu >0$, while the dashed lines correspond to $\mu <0$.\\  The upper two lines
in the vicinity of
$m_0=100$ are for $m_{1/2}=100$ GeV, and the lower two lines are for
$m_{1/2}=150$ GeV.\\
$\lambda_{{{t}_G}}=3.54 $ for all the lines}.
\item[Fig. 1(b)~:] {$d_r$ for  model (2)  is
plotted as a function of of the universal soft mass $m_0$.\\ The solid lines
correspond to
$\mu >0$ , while the dashed lines correspond to $\mu <0$.\\ The upper two lines
in the vicinity of $m_0=150$ GeV are for $m_{1/2}=160$ GeV, and the lower two lines
are for
$m_{1/2}=200$ GeV.\\ $\lambda_{{{t}_G}}=3.54 $ for all the lines}.
\item[Fig. 1(c)~:] {$d_r$ for Model (3) is plotted as a function of the
universal soft SUSY breaking gaugino mass $m_{1/2}$.\\ The solid lines
correspond
to
$\mu>0$, and the dashed lines correspond to $\mu<0$.\\The upper two lines around
$m_0=150$ GeV are for
$m_{1/2}=190$ GeV, and the lower two lines in that region are for $m_{1/2}=220$
GeV.\\ $\lambda_{{{t}_G}}=3.54 $ for all the lines.}

\item[Fig. 1(d)~:] {$d_r$ for Model (4) is plotted as a function of
$\log_{10}{M_I/GeV}$.\\ The solid lines correspond to $\mu>0$, and the dashed
lines correspond to
$\mu<0$.\\ The upper two lines around $M_I=10^8$ GeV correspond to
$\lambda_{{t}_G} =3.54$, and the lower two lines in the same region correspond
to
$\lambda_{{{t}_G}}=1.38 $. $m_0$ = $m_{1/2}=180$ GeV for all the lines. }

\item[Fig. 2(a)~:] {The $m_b^{\rm pole}$ values are plotted as a function of
$\log_{10}{M_I/{\rm GeV}}$ in model(4) with complete third generation Yukawa
coupling unification. The solid line corresponds to $\alpha_s=0.117$,
sin$^2 \theta_W$=.2332 (within 2-$\sigma$ of the experimental mid-value) and
$\alpha
$=1/127.9 at the $M_Z$ scale, the dashed line corresponds to
$\alpha_s=0.122$, sin$^2 \theta_W$=.2321 (the experimental mid-value) and
$\alpha
$=1/127.9 at the $M_Z$ scale. \\ $\lambda_{{{t}_G}}=3.54 $ for both of the
lines.}

\item[Fig. 2(b)~:] {The $m_b^{\rm pole}$ values are plotted as a function of
$\tan\beta$ in model(4) with $M_I=10^{12}$ GeV. The solid and dashed lines
correspond to the same values of $\alpha_s$ as in Fig. 2(a). \\
$\lambda_{{{t}_G}}=3.54 $ for both of the lines.}
\end{itemize}
\newpage
\begin{figure}[htb]
\centerline{ \DESepsf(figs15.epsf width 15 cm) } \smallskip
\nonumber
\end{figure}
\vfill
\begin{figure}[htb]
\centerline{ \DESepsf(figs2a5.epsf width 15 cm) } \smallskip
\nonumber
\end{figure}
\vfill
\begin{figure}[htb]
\centerline{ \DESepsf(figs2b5.epsf width 15 cm) } \smallskip
\nonumber
\end{figure}

\newpage
\begin{thebibliography}{[001]}

\bibitem{[SS]}R. Shafer and Q. Shafi, Nature (London) {\bf 359}, 199
(1992).

\bibitem{[axion]}M. Dine, W. Fischler and  M. Srednicki, Phys. Lett.
{\bf B104}, 199 (1981).

\bibitem{[ddk]}N. G. Deshpande, B. Dutta, and E. Keith, hep-ph/9512398 (to
appear in Phys. Rev. {\bf D} as a Rapid Comm.).

\bibitem{[dh]} S. Dimopoulos and L. J. Hall, Phys. Lett. {\bf B344}, 185
(1995).

\bibitem{[BHS]} R. Barbieri, L. J. Hall, and A. Strumia,  Nucl. Phys.
{\bf B449}, 437  (1995).

\bibitem{[AS]} R. Barbieri, L. J. Hall, and A. Strumia,  Nucl. Phys.
{\bf B445}, 219  (1995).

\bibitem{[ML10]}F. Vissani and A. Y. Smirnov, Phys. Lett. \underline{B341} 173
(1994); A. Brignole, H. Murayama, and R. Rattazzi,  Physics Lett.
{\bf B335}, 345  (1994);

\bibitem{[ddku]}N. G. Deshpande, B. Dutta and E. Keith, hep-ph/9604236.

\bibitem{[MohaLee]}D-G. Lee and R. N. Mohapatra, Phys. Rev. {\bf D52}, 4125
(1995).

\bibitem{[EMa]}E. Ma, Phys. Rev. {\bf D51}, 236  (1995).

\bibitem{[DKR]}N. G. Deshpande, E. Keith, and T. G. Rizzo, Phys. Rev. Lett. {\bf
70}, 3189 (1993).

\bibitem{[dn]} Altarev et.al., Phys. Lett. {\bf B276}, 242 (1992) .

\bibitem{[de]} Commins et.al.,  LBL-35572.
\bibitem{[SUSYYUK]}For example, see: V. Barger, M. Berger, and P. Ohmann,
Phys. Rev.
{\bf D47}, 1093 (1993); V. Barger, M. Berger, and P. Ohmann, Phys. Rev.
{\bf D49}, 4908 (1994).

\bibitem{[mb]} M. Neubert, SLAC preprint SLAC-PUB-6263 (1993) and references
therein.

\bibitem{[BM]}B. Brahmachari and R. N. Mohapatra, Phys. Lett.
{\bf B357}, 566 (1995).

\end{thebibliography}
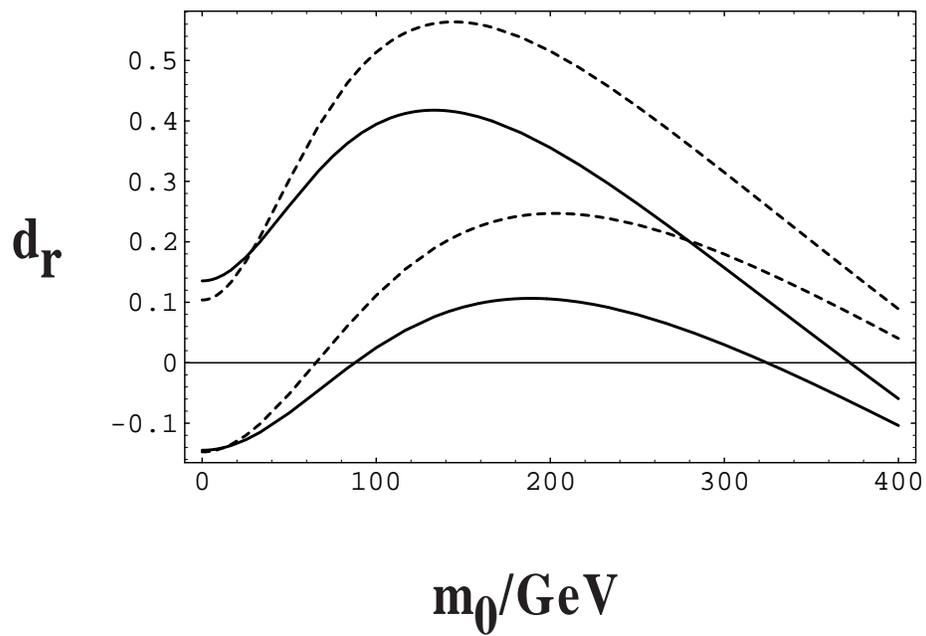
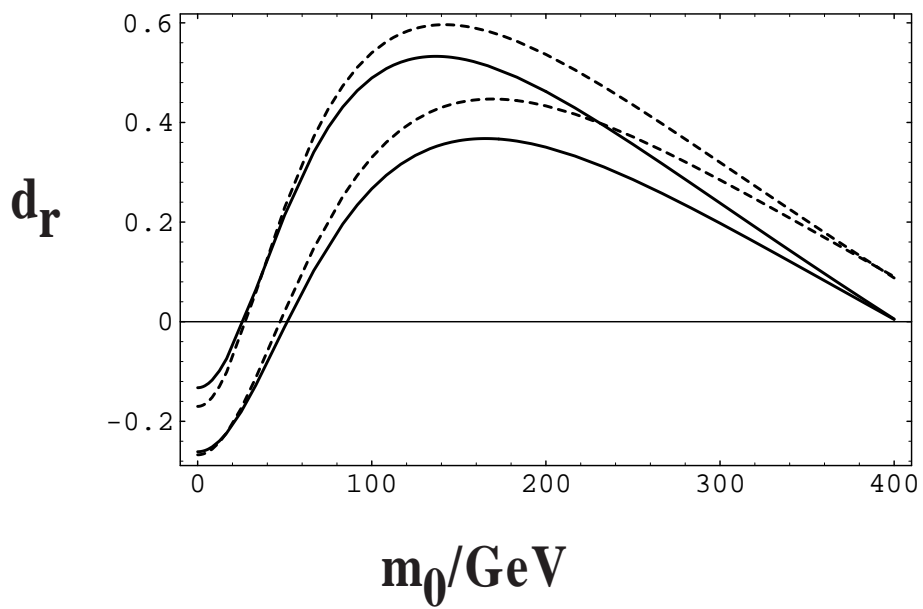
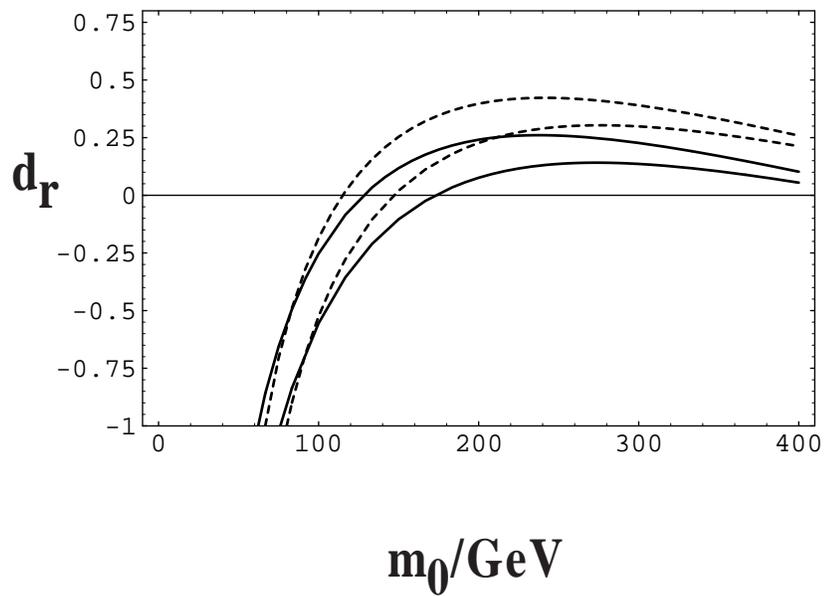
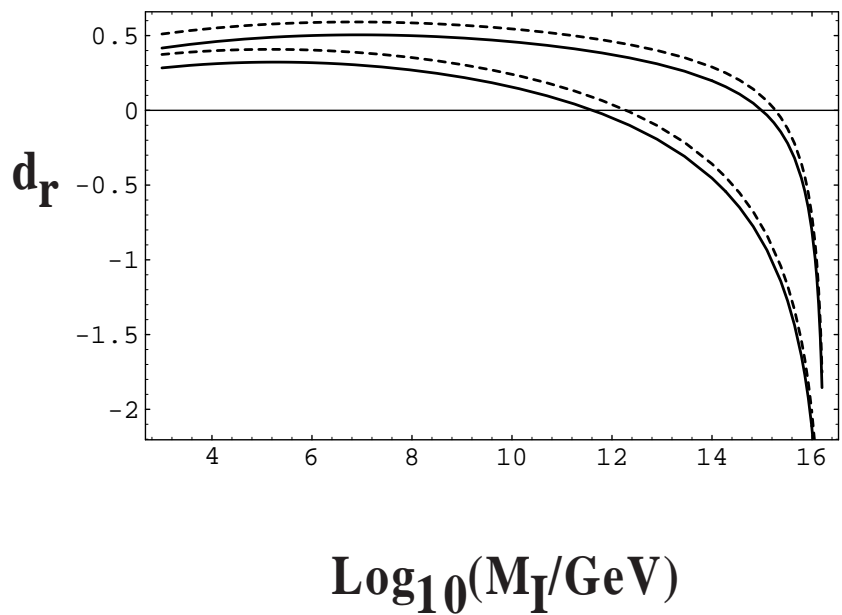
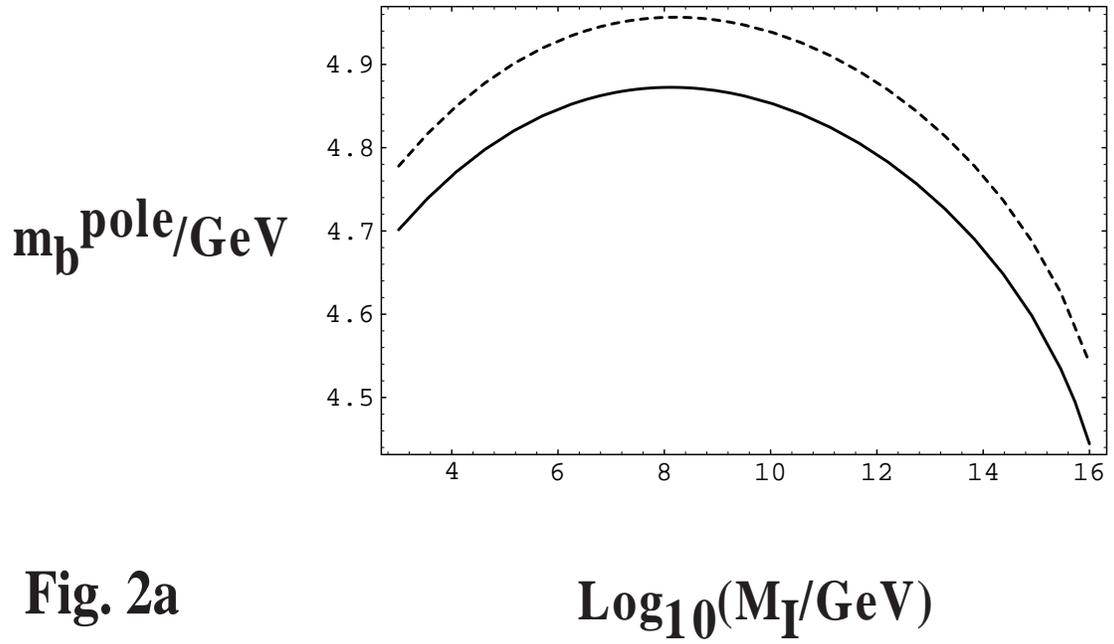
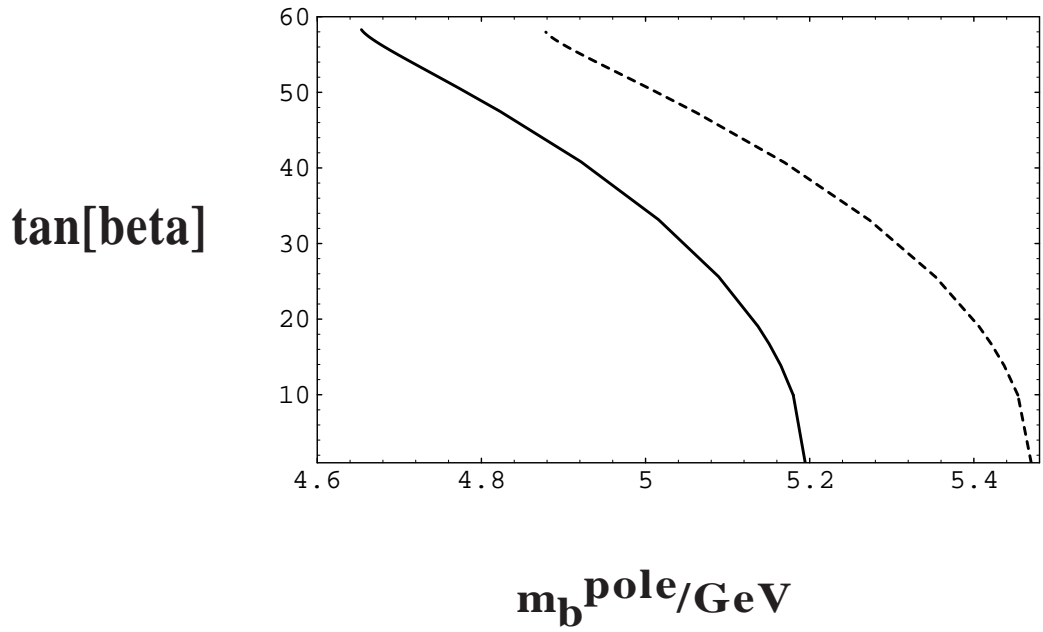
\end{document}